\documentclass{ws-procs975x65}

\begin{document}



\title{3D RELATIVISTIC MHD SIMULATION OF A TILTED ACCRETION DISK AROUND A RAPIDLY ROTATING BLACK HOLE\footnote{This research has been partially supported by
a REAP grant from the South Carolina Space Grant Consortium.}}

\author{P. CHRIS FRAGILE}

\address{Dept of Physics \& Astronomy, College of Charleston, Charleston, SC 29424 USA\\
\email{fragilep@cofc.edu}}

\author{PETER ANNINOS}

\address{Lawrence Livermore National Laboratory, Livermore, CA 94550 USA\\
}

\author{OMER M. BLAES}

\address{University of California, Santa Barbara, CA 93106 USA\\
}

\author{JAY D. SALMONSON}

\address{Lawrence Livermore National Laboratory, Livermore, CA 94550 USA\\
}


\begin{abstract}
We posit that accreting compact objects, including stellar mass
black holes and neutron stars as well as supermassive black holes,
may undergo extended periods of accretion during which the angular
momentum of the disk at large scales is misaligned with that of the
compact object. In such a scenario, Lense-Thirring precession caused
by the rotating compact object can dramatically affect the disk. In
this presentation we describe results from a three-dimensional
relativistic magnetohydrodynamic simulation of an MRI turbulent disk
accreting onto a tilted rapidly rotating black hole. For this case,
the disk does not achieve the commonly described Bardeen-Petterson
configuration; rather, it remains nearly planar, undergoing a slow
global precession. Accretion from the disk onto the hole occurs
predominantly through two opposing plunging streams that start from
high latitudes with respect to both the black-hole and disk
midplanes. This is a consequence of the non-sphericity of the
gravitational spacetime of the black hole.
\end{abstract}

\bodymatter

\section{Introduction}\label{intro}
Recently, our group has been working on three-dimensional, fully
relativistic simulations of tilted thick disks, first in the
hydrodynamic limit \cite{fra05b} and here in full general
relativistic MHD (GRMHD). A general relativistic treatment ensures
that all important relativistic features, such as the cusp in the
potential and the Einstein and Lense-Thirring precessions of the
orbits, as well as any higher order couplings of these features, are
treated properly.
The inclusion of magnetic fields is important because it is now
widely believed that local ``viscous'' stresses are generated by
turbulence that results from the magnetorotational instability (MRI)
\cite{bal91}.


This work is carried out using the Cosmos++ astrophysical
magnetohydrodynamics code \cite{ann05}. For this work, the key
features of Cosmos++ are the flexible mesh structure, adaptive
gridding, and three-dimensional general relativistic MHD
capabilities. This simulation is carried out on a spherical-polar
mesh with an equivalent resolution of $64^3$ zones. The full set up
of this problem and higher resolution simulations will be described
in a forthcoming paper (Fragile et al., in preparation).

\section{Results}
\label{sec:results}

\begin{figure}
\begin{center}
\includegraphics[scale=0.4]{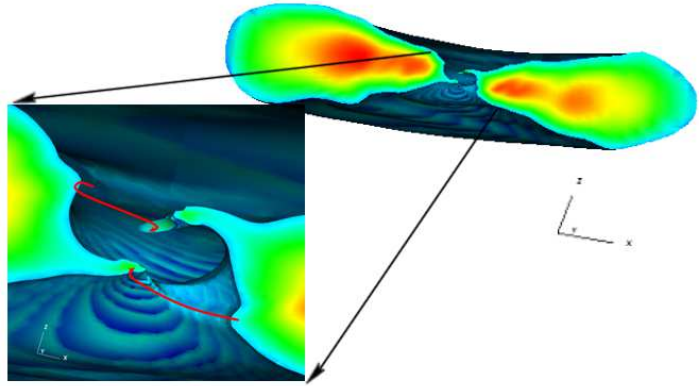}
\end{center}
\caption{Logarithmic density plot from the final data dump ($t=10$
orbits at the pressure maximum of the initial torus). This figure is
oriented such that the black-hole spin axis points straight up. The
$z$-axis of the triad gives the original orientation of the disk
angular momentum vector. The main body of the disk shows very little
warping or realignment toward the symmetry plane of the black hole.
The plunging region (shown in inset) shows two opposing,
high-latitude streams of material connecting the disk to the horizon
(indicated by arrows). \label{fig:plunge}}
\end{figure}

Perhaps the most striking feature of the simulated disk at late
times are the two opposing plunging streams that start from high
latitudes both with respect to the black-hole symmetry plane and the
disk midplane. Figure \ref{fig:plunge} shows the structure of the
full disk and a zoomed in view of the plunging region with the
plunging streams. Note that one stream remains entirely above the
black-hole symmetry plane, while the other remains below. As
material passes through this plunging stream it undergoes strong
differential precession. The precession totals approximately
$180^\circ$; thus, the material enters the black hole from the
opposite azimuth from which it began its plunge.

\begin{figure}[ht]
    \hfill
    \begin{minipage}[t]{.47\textwidth}
     \begin{center}
      \epsfig{file=isco2.eps, scale=0.25}
      \caption{Plot of the inclination dependence of $r_{ISCO}$
for black-hole spins $a=0$, 0.5, 0.9, and 0.998. Inclinations $0 \le
i \le 90^\circ$ represent prograde orbits, whereas inclinations
$90^\circ \le i \le 180^\circ$ represent retrograde orbits.}
      \label{fig:isco}
     \end{center}
    \end{minipage}
    \hfill
    \begin{minipage}[t]{.47\textwidth}
     \begin{center}
      \epsfig{file=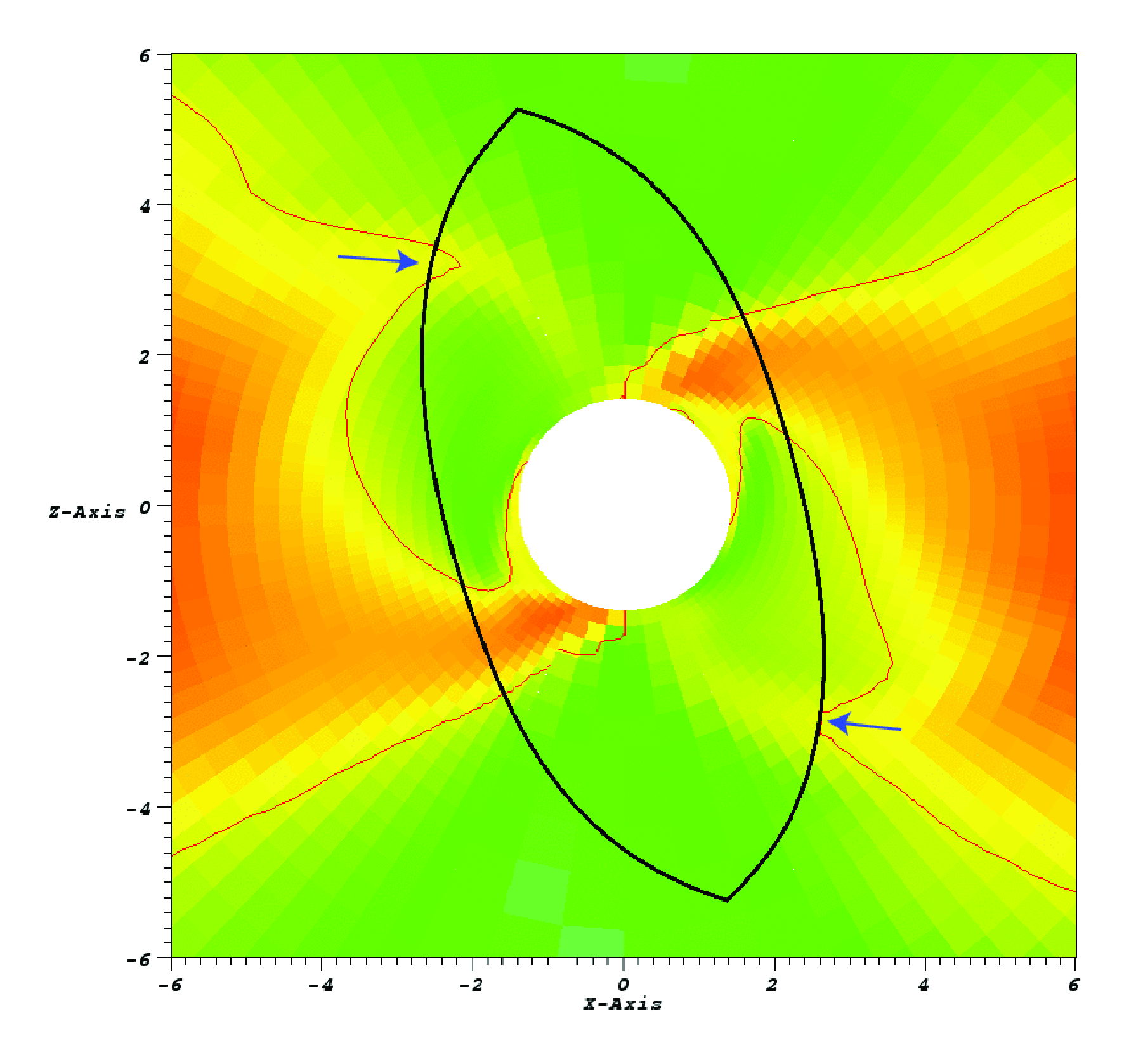, scale=0.18}
      \caption{Meridional plot ($\varphi=0$) through the final dump of the
simulation showing the logarithm of density overlaid with a polar
plot of the ``ISCO surface'' for an $a=0.9$ black hole. Notice that
the plunging streams start near the largest cylindrical radius
($r\cos\vartheta$) of this surface (indicated by arrows).}
      \label{fig:slice}
    \end{center}
    \end{minipage}
    \hfill
\end{figure}

The interesting question from Figure \ref{fig:plunge} is {\em why}
do these opposing accretion streams start from such high latitude
with respect to the black-hole symmetry plane and disk midplane?
Fundamentally this is due to the aspherical nature of the
gravitational spacetime of the rotating black hole. A nice way to
illustrate this is to consider the dependence of $r_{ISCO}$, the
coordinate radius of the innermost stable circular orbit, on
inclination. Figure \ref{fig:isco} illustrates this dependence for a
few selected cases of $a$. The key point of the plot is that orbital
stability around a rotating black hole is strongly dependent on the
inclination of the orbit. Notice that the unstable region increases
monotonically for increasing inclination.

We can make better use of the information in Figure \ref{fig:isco}
by converting it to a polar plot and overlaying it on a plot of data
from the simulation, as is done in Figure \ref{fig:slice}. The polar
plot creates an ``ISCO surface'', which gives a clear indication of
where the most unstable regions are. Note that the plunging orbits
highlighted previously start near where the disk first encounters
the ISCO surface. More precisely, the streams start near the largest
cylindrical radius ($r\cos\vartheta$) of the ISCO surface, measured
with respect to the angular momentum axis of the disk. The increase
in $r_{ISCO}$ with inclination then explains why these plunging
orbits start at high inclinations relative to the black-hole
symmetry plane and the disk midplane, since this is where the
gravitational potential is most unstable.

\section*{Acknowledgments}
This work was supported under the following NSF programs:
Partnerships for Advanced Computational Infrastructure, Distributed
Terascale Facility (DTF) and Terascale Extensions: Enhancements to
the Extensible Terascale Facility.

\vfill

\end{document}